\documentclass[superscriptaddress,showpacs,preprintnumbers,twocolumn,nofootinbib]{revtex4}

\usepackage{amsmath,amssymb}
\usepackage{graphicx}

\newcommand{\mbf}[1]{\mathbf{#1}}

\begin{document}
\preprint{DO-TH 11/09}

\title{Family symmetries and alignment in multi-Higgs doublet models}
\author{Ivo de Medeiros Varzielas}
\email{ivo.de@udo.edu}
\affiliation{Fakult\"{a}t f\"{u}r Physik, Technische Universit\"{a}t Dortmund D-44221 Dortmund, Germany}

%11.30.Hv Family symmetries
%12.15.Ff Fermion Masses and Mixing
%12.60.Fr Multi Higgs models
%Maybe also 14.80.Cp Non-standard Higgs

\begin{abstract}
The exact alignment of the Yukawa structures on multi-Higgs doublet models provides cancellation of tree-level flavour changing couplings of neutral scalar fields. We show that family symmetries can provide a suitable justification for the Yukawa alignment.
\end{abstract}
\maketitle

\section{Introduction \label{Intro}}

A Multi-Higgs Doublet Model (MHDM) consists in a straightforward generalisation of the Standard Model (SM) where extra $SU(2)$ doublet scalars are added to the field content. With just one extra doublet added, the two-Higgs doublet model is a particular case of the MHDM \cite{Haber, Gustavo}.

In the MHDM there are Yukawa couplings associated to each Higgs doublet for each family of fermions - up quarks, down quarks, charged leptons and neutrinos if Right-Handed (RH) neutrinos are also added. It is well known that this presents the potential for large unobserved flavour changing processes such as Flavour Changing Neutral Currents (FCNCs). One can see this in the Higgs basis in which the mass matrices of the fermions come from only those Yukawa matrices associated with a particular doublet. For a given family, that mass matrix can be diagonalised - but without further assumption the other Yukawa matrices of the family are arbitrary complex matrices which would enable large tree-level FCNCs.
It has been noted \cite{PT_a} that these unobserved processes completely cancel in the exact alignment limit i.e. all Yukawa matrices of a given family are perfectly aligned. It was shown \cite{Lavoura_RG} that such alignment can not be preserved by renormalisation unless additional symmetries are imposed, although the contributions to the unobserved processes due to this misalignment can be compatible with the current experimental constraints for regions of the parameter space \cite{Ibarra_RG}.

It is reasonable to expect that problematic processes will be suppressed when there is approximate alignment. In analogy to the solution of the SUSY flavour problem where Family Symmetries (FSs) provide approximate alignment of fermions and sfermions (see e.g. \cite{Oscar1, Oscar2, Graham}), a suitable FS is a good candidate solution to address the MHDM flavour problem by providing approximate alignment of the Yukawas of each $SU(2)$ doublet. As in the FS solution to the SUSY flavour problem, this would be particularly appealing if the FS solves the MHDM issues while simultaneously addressing the flavour problems of the SM (such as the otherwise unexplained hierarchy in fermion masses).

In the most extreme cases, the FS can provide perfect alignment and protect it from renormalisation effects.
Here we present models where the FS is used solely to address the potential flavour problems of the MHDM by achieving perfect alignment for the given families - without attempting to ameliorate the flavour issues of the SM.

The exact Yukawa alignment in the MHDM is achieved through a specific strategy that combines two requirements: only one FS Invariant Combination (FSIC) is allowed for each family; all the Higgs $SU(2)$ doublets are singlets of the FS, such that the single allowed FSIC can be made invariant under the SM through coupling to any of the Higgs doublets. We then argue that as a generalisation of this strategy, dropping the constraining single FSIC requirement while maintaining the Higgs as singlets of the FS is a promising approach to achieve approximate Yukawa alignment.

\section{Simple alignment example}

In order to illustrate the proposed strategy we use $SU(3)_{[]} \otimes SU(3)_{()} \otimes C_n$ as the FS. The fermions are assigned as triplets. To allow a FSIC, familons (SM singlet scalars) are added and assigned as anti-triplets under the FS.
The requirement that each family has a single allowed coupling is simple conceptually, but it can be quite difficult to implement. In order to do so, an auxiliary Abelian factor ($C_n$) is not enough. This is why we resort to two distinct $SU(3)$ (c.f. \cite{Ivo1, Ivo2}) as with two non-Abelian factors the Left-Handed (LH) sectors ($Q$ and $L$) are separated from the RH sectors ($u^c$, $d^c$, $e^c$ and $\nu^c$). $C_n$ is then sufficient to keep both $Q$ separate from $L$ and each RH sector separate from one another.

We start with the goal of Yukawa alignment for a single family, e.g. the up-type quarks. $C_n$ is not required if one only wants to obtain alignment for a single family, as with only two familons there is only one FSIC: $[\phi_Q^i Q_i] (\phi_u^j u^c_j)$. It is made invariant under the SM by coupling to any of the $N$ Higgs doublets $H_A$:
\begin{equation}
\label{eq:SU3Lup}
\mathcal{L}_{u} =  \sum^N_{A=1} c^u_A H^\dagger_A [\phi_Q^i Q_i] (\phi_u^j u^c_j) + h.c.
\end{equation}
The SM invariant contractions are implicitly assumed in this compact notation. The square brackets denote invariant contractions under $SU(3)_{[]}$ and the brackets denote invariant contractions under $SU(3)_{()}$, with the generation indices $i,j$ used with superscript for anti-triplets and subscript for triplets. The distinct familons have a subscript label (this is simply a notational label and not an index). $c^u_A$ is the arbitrary coupling with the superscript label denoting the family and the subscript label denoting the Higgs doublet. $\mathcal{L}_{u}$ is non-renormalisable and we do not display explicitly the necessary messenger mass scales associated with the UV completion of the model. An explicit UV completion such as those presented in \cite{IvoLuca} is beyond the scope of the present work, but we note that in principle the completions should not affect the main results as the $H_A$ are FS singlets.

In this implementation, the non-Abelian symmetries are required to keep the LH and RH separate - otherwise $\sum^N_{A=1} c'^u_A H_A^{\dagger} (\phi_u^i Q_i) (\phi_Q^j u^c_j)$ would be added to the single invariant term shown in eq.(\ref{eq:SU3Lup}) and invalidate the rather constraining strategy we are implementing. It is relevant to note how this contrasts with \cite{Ivo1, Ivo2} where a single $SU(3)$ is used and both terms where the same familon couples to the LH and to the RH are explicitly needed to obtain the desired phenomenology - these two approaches are not compatible.

The next goal is going to the full quark sector. It may be interesting to require alignment only for the quarks as this type of strategy can be embedded into a leptophobic (or lepton-inert) MHDM. To keep only a single FSIC for the ups and another for the downs, the quark RH sectors need to be distinguished. $C_n$ with $n=2$ is sufficient to keep $\phi_u$, $u^c$ separate from $\phi_d$, $d^c$ by charging e.g. the latter two fields non-trivially. The SM-invariants are then built by adding the $H_A$:
\begin{equation*}
\mathcal{L}_{Q} =  \sum^N_{A=1} [\phi_Q^i Q_i] \left( c^d_A H_A (\phi_d^j d^c_j) + c^u_A H^\dagger_A (\phi_u^j u^c_j) \right) + h.c.
\end{equation*}

When both leptons and quarks are considered, $C_n$ must also keep the $\phi_Q$, $Q$ fields from interfering with the $\phi_L$, $L$ fields.
Trying the same $C_2$ that worked for quarks, $\phi_L$ would need to transform non-trivially to distinguish it from $\phi_Q$ and that inevitably allows FSICs like $[\phi_L Q] (\phi_u d^c)$. In this $SU(3)_{[]} \otimes SU(3)_{()} \otimes C_n$ framework we found $n=7$ to be the smallest $n$ that works when charged leptons are considered for alignment. If Yukawa alignment is imposed additionally to neutrinos, $C_{10}$ can be used. Possible assignments are listed in Table \ref{ta:SU3C7} for $n=7$ and Table \ref{ta:SU3C10} for $n=10$, where $\alpha^n = 1$.

\begin{table}[t]
\begin{tabular}{|c|ccc|cc|ccc|cc|}
\hline
	& $Q$ & $u^c$ & $d^c$ & $L$ & $e^c$ & $\phi_Q$ & $\phi_u$ & $\phi_d$ & $\phi_L$ & $\phi_e$ \\
\hline
$SU(3)_{[]}$ & $\mbf{3}$ & $\mbf{1}$ & $\mbf{1}$ & $\mbf{3}$ & $\mbf{1}$ & $\mbf{\bar{3}}$ & $\mbf{1}$ & $\mbf{1}$ & $\mbf{\bar{3}}$ & $\mbf{1}$ \\
$SU(3)_{()}$ & $\mbf{1}$ & $\mbf{3}$ & $\mbf{3}$ & $\mbf{1}$ & $\mbf{3}$ & $\mbf{1}$ & $\mbf{\bar{3}}$ & $\mbf{\bar{3}}$ & $\mbf{1}$ & $\mbf{\bar{3}}$ \\
$C_7$ & 1 & 1 & $\alpha^3$ & $\alpha$ & $\alpha^4$ & $\alpha$ & $\alpha^6$ & $\alpha^3$ & $\alpha^2$ & 1 \\
\hline
\end{tabular} 
\caption{\label{ta:SU3C7} $SU(3)_{[]} \otimes SU(3)_{()} \otimes C_7$ assignments. $\alpha^7 = 1$.}
\end{table}

\begin{table}[t]
\begin{tabular}{|c|ccc|ccc|ccc|ccc|}
\hline
	& $Q$ & $u^c$ & $d^c$ & $L$ & $e^c$ & $\nu^c$ & $\phi_Q$ & $\phi_u$ & $\phi_d$ & $\phi_L$ & $\phi_e$ & $\phi_\nu$ \\
\hline 
$SU(3)_{[]}$ & $\mbf{3}$ & $\mbf{1}$ & $\mbf{1}$ & $\mbf{3}$ & $\mbf{1}$ & $\mbf{1}$ & $\mbf{\bar{3}}$ & $\mbf{1}$ & $\mbf{1}$ & $\mbf{\bar{3}}$ & $\mbf{1}$ & $\mbf{1}$ \\
$SU(3)_{()}$ & $\mbf{1}$ & $\mbf{3}$ & $\mbf{3}$ & $\mbf{1}$ & $\mbf{3}$ & $\mbf{3}$ & $\mbf{1}$ & $\mbf{\bar{3}}$ & $\mbf{\bar{3}}$ & $\mbf{1}$ & $\mbf{\bar{3}}$ & $\mbf{\bar{3}}$ \\
$C_{10}$ & 1 & 1 & $\alpha^3$ & $\alpha$ & $\alpha^4$ & $\alpha^7$ & $\alpha$ & $\alpha^9$ & $\alpha^6$ & $\alpha^2$ & $\alpha^3$ & 1 \\
\hline
\end{tabular} 
\caption{\label{ta:SU3C10} $SU(3)_{[]} \otimes SU(3)_{()} \otimes C_{10}$ assignments. $\alpha^{10} = 1$.}
\end{table}

With the symmetries assignments of Table \ref{ta:SU3C7} or Table \ref{ta:SU3C10} we have the straightforward extension of eq.(\ref{eq:SU3Lup}):
\begin{eqnarray*}
\mathcal{L} = \sum^N_{A=1} \left( c^d_A H_A [\phi_Q^i Q_i] (\phi_d^j d^c_j) + c^u_A H^\dagger_A [\phi_Q^i Q_i] (\phi_u^j u^c_j) + \right. \\
\left.  c^e_A H_A [\phi_L^i L_i] (\phi_e^j e^c_j) + c^\nu_A H^\dagger_A [\phi_L^i L_i] (\phi_\nu^j \nu^c_j) \right) + h.c.
\end{eqnarray*}
The term with $\nu^c$ should be omitted if the particle content is the one from Table \ref{ta:SU3C7}. Considering Majorana neutrino masses and types of see-saw is beyond the scope of the present work.

After the familons acquire Vacuum Expectation Values (VEVs) $\langle \phi \rangle$, the Yukawa couplings for family $f$ and for $H_A$ are $3 \times 3$ matrices in generation space that depend only on $c^f_A$ and on the respective VEVs:
\begin{equation}
Y_{f A}^{ij} =c^f_A \langle \phi_F^i \rangle \langle \phi_f^j \rangle \,.
\end{equation}
Each given family has only the exact same FSIC for each $H_A$ (with an arbitrary coupling constant $c^f_A$), consequently the Yukawa matrices feature the same familon VEVs and are therefore aligned.
The proportionality coefficients between Yukawa matrices of the same family but different $H_A$ are given by ratios of the respective $c^f_A$ couplings. In a basis where one of the Yukawa matrices of that family is diagonal, all other Yukawa matrices of that family are simultaneously diagonalised. The alignment hypothesis \cite{PT_a} is effectively implemented by the FS.

We have not used the FS to address any flavour issues of the SM - in fact these models feature Yukawa matrices with two vanishing eigenvalues even with the most general familon VEVs. Before we consider a case where this is avoided and the FS implements exact Yukawa alignment, it is interesting to discuss in more detail the single FSIC requirement. If we lift this constraint while the Higgs doublets are kept FS singlets, all FSICs can be made SM invariant by coupling to any of the $H_A$. It was the single FSIC requirement that made the strategy incompatible with the approach in e.g. \cite{Ivo1}, where extra FSICs are used to explain the fermion masses. In such an approach it is natural to expect some hierarchy between a dominant FSIC and additional FSICs (through an hierarchy of the familon VEVs or by having additional familon insertions). By embedding a MHDM into that approach, it would have Yukawa alignment at leading order, the alignment being affected by the extra terms only at higher orders. If the extra terms are sufficiently suppressed compared to the leading order FSIC, the approximate alignment resulting from the FS could be sufficient to avoid problematic FCNCs.

\section{Another alignment model}

The strategy of combining a single FSIC with FS singlets $H_A$ can be implemented without leading to Yukawa matrices with zero eigenvalues. We choose now the FS $PSL_2(7) \otimes C_n$. The group $PSL_2 (7)$ is a discrete subgroup of $SU(3)$ \cite{Luhn1} that is particularly interesting for embedding into $SO(10)$ GUTs \cite{Steve1, Steve2} (although we do not attempt such an embedding). We use familons $\chi$ that transform as sextuplets of the group, while the fermions $F$ (LH), $f^c$ (RH) are triplets and the Higgs $SU(2)$ doublets are singlets of the FS. In order to illustrate the FSICs we first require alignment for a single family:
\begin{equation}
\label{eq:Lf}
L_{f} =  \sum^N_{A=1} c^f_A H_A (F_j \hat{\chi}_f^{jl} f^c_l) \,.
\end{equation}
The SM invariant is again implicit and $j,l$ are the generation indices (every other superscripts or subscripts are notational labels). The brackets and $\hat{\chi}_f$ denote the $PSL_2 (7)$ invariant constructed out of the sextuplet and the two triplets \cite{Luhn1, Steve1, Steve2}:
\begin{equation}
\hat{\chi}_f^{jl} = -\frac{(1+\mathrm{i})}{6 \sqrt{2}} \sum^6_{r=1} \left[ \chi_f^r S_r^{jl}\right] \,,
\end{equation}
where we denote the six components of the sextuplet $\chi_f$ as $\chi_f^r$. The six matrices $S_r^{jl}$ are $3 \times 3$ matrices (notice the generation indices $j,l$). They encode the tensorial product that makes the $PSL_2 (7)$ invariant of a sextuplet with two triplets, as in eq.(\ref{eq:Lf}). The matrices in a given convenient basis are \cite{Luhn1, Steve1, Steve2}:
\begin{equation*}
S_1 = 
\left( \begin{array}{ccc}
	4 & 1 & 1\\
	1 & -2 & -2\\
	1 & -2 & -2
\end{array} \right) \, ;
\, S_2 = - \mathrm{i} \sqrt{3}
\left( \begin{array}{ccc}
	0 & 1 & -1\\
	1 & 2 & 0\\
	-1 & 0 & -2
      \end{array} \right) \, ;
\end{equation*}
\begin{equation*}
S_3 = - \mathrm{i} \sqrt{3} \, b_7
\left( \begin{array}{ccc}
	0 & 1 & -1\\
	1 & -1 & 0\\
	-1 & 0 & 1
\end{array} \right) \, ;
\end{equation*}
\begin{equation*}
S_4 = \sqrt{3} \, b_7
\left( \begin{array}{ccc}
	0 & 1 & 1\\
	1 & 1 & 0\\
	1 & 0 & 1
      \end{array} \right) \,;
\end{equation*}
\begin{equation*}
S_5 = \sqrt{2}
\left( \begin{array}{ccc}
	2 & -1 & -1\\
	-1 & 2 & -1\\
	-1 & -1 & 2
\end{array} \right) \, ;
\, S_6 = - \mathrm{i} \sqrt{6} \, \bar{b}_7
\left( \begin{array}{ccc}
	1 & 0 & 0\\
	0 & 0 & 1\\
	0 & 1 & 0
      \end{array} \right) \,,
\end{equation*}
where $b_7 = \frac{1}{2} \left(-1 + \mathrm{i}\sqrt{7} \right)$ and $\bar{b}_7$ is its complex conjugate.

In order to provide Yukawa alignment through having a single FSIC for each family we use the sextuplet familons $\chi_u$, $\chi_d$, $\chi_e$ and $\chi_\nu$. We have found that $C_7$ is sufficient to preserve this strategy at the level of cubic FSICs. Possible assignments are listed in Table \ref{ta:PSL_C7}.

\begin{table}[t]
\begin{tabular}{|c|ccc|ccc|cc|cc|}
\hline
	& $Q$ & $u^c$ & $d^c$ & $L$ & $e^c$ & $\nu^c$ & $\chi_u$ & $\chi_d$ & $\chi_e$ & $\chi_\nu$ \\
\hline
$PSL_2(7)$ & $\mbf{3}$ & $\mbf{3}$ & $\mbf{3}$ & $\mbf{3}$ & $\mbf{3}$ & $\mbf{3}$ & $\mbf{6}$ & $\mbf{6}$ & $\mbf{6}$ & $\mbf{6}$ \\
$C_7$ & $\alpha$ & $\alpha^4$ & $\alpha^2$ & $\alpha^2$ & 1 & $\alpha^5$ & $\alpha^2$ & $\alpha^4$ & $\alpha^5$ & 1 \\
\hline
\end{tabular}
\caption{\label{ta:PSL_C7} $PSL_2 (7) \otimes C_7$ assignments. $\alpha^7 = 1$.}
\end{table}

The Yukawa Lagrangian terms (one for each family) are of the type of eq.(\ref{eq:Lf}). They are non-renormalisable and therefore higher order combinations of the triplets and sextuplets may contribute to the effective Yukawa couplings and invalidate the strategy we are trying to implement. In terms of quartic FSICs, $PSL_2(7)$ allows two distinct combinations to be constructed from two triplets and two sextuplets, so terms like $L e^c \chi_e \chi_\nu$ would appear with the assignments of Table \ref{ta:PSL_C7}. One possible solution to this drawback lies in developing a suitable renormalisable UV completion. As shown in \cite{IvoLuca} the requirement of renormalisability together with an appropriate messenger content can strongly limit the allowed invariants. Terms that one would otherwise expect to be present at the level of building the non-renormalisable effective theory are actually absent in the effective theory limit associated with the UV complete model.

A suitable UV completion is outside the scope of the work we present here, so instead we propose augmenting the $C_n$ symmetry in order to forbid terms up to a given order.
At the effective level it is not possible to have $C_n$ forbid extra terms up to arbitrarily high order. For example, two sextuplets can combine into a $PSL_2 (7)$ singlet: $\left((\chi_u)^{n_1} (\chi_d)^{n_2} (\chi_e)^{n_3} (\chi_\nu)^{n_4} \right)^n$ will always be a complete $PSL_2(7) \otimes C_n$ singlet regardless of the charge assignments and for any choice of non-negative integers $n_1, n_2, n_3, n_4$ that corresponds to an even number of sextuplet insertions. Any such combination can be appended to any of the desired cubic FSIC and produce extra FSICs. This particular example clearly demonstrates that suppressing to arbitrarily high order is not worth looking for. It is also extremely likely that there would be accidental terms of different types at orders much lower that depend on the charge assignments (such as the $L e^c \chi_e \chi_\nu$ in the $C_7$ model).
In any case if one decides to prevent terms up to the quartic level of FSICs, this can be achieved by the $C_{18}$ assignments in Table \ref{ta:PSL_C18}.

\begin{table}[t]
\begin{tabular}{|c|ccc|ccc|cc|cc|}
\hline
	& $Q$ & $u^c$ & $d^c$ & $L$ & $e^c$ & $\nu^c$ & $\chi_u$ & $\chi_d$ & $\chi_e$ & $\chi_\nu$ \\
\hline
$PSL_2(7)$ & $\mbf{3}$ & $\mbf{3}$ & $\mbf{3}$ & $\mbf{3}$ & $\mbf{3}$ & $\mbf{3}$ & $\mbf{6}$ & $\mbf{6}$ & $\mbf{6}$ & $\mbf{6}$ \\
$C_{18}$ & 1 & $\alpha^{10}$ & $\alpha^{9}$ & $\alpha^2$ & $\alpha^{6}$ & $\alpha^{5}$ & $\alpha^{8}$ & $\alpha^{9}$ & $\alpha^{10}$ & $\alpha^{11}$ \\
\hline
\end{tabular}
\caption{\label{ta:PSL_C18} $PSL_2 (7) \otimes C_{18}$ assignments. $\alpha^{18} = 1$.}
\end{table}

Regardless if it is achieved through a suitable UV completion or another mechanism, we assume extra FSICs are not allowed (or extremely suppressed). If so, after the familons acquire VEVS the Yukawas matrices depend only on the respective coupling $c^f_A$ and VEVs
\begin{equation}
Y_{f A}^{jl} =c^f_A \left( \sum_{r=1}^6 \langle \chi_f^{r} \rangle S_{r}^{jl} \right).
\end{equation}
For a given family $f$ these $3 \times 3$ matrices have the exact same structure for any $H_A$.

Finally, we do not consider RH neutrinos masses here as the topic we concern ourselves with is the alignment of Yukawa couplings. The allowed terms depend on the $C_n$ assignments - for example those of Table \ref{ta:PSL_C7} allow the term $\chi_d \nu^c \nu^c$ whereas those on Table \ref{ta:PSL_C18} would allow $\chi_u \nu^c \nu^c$, but other assignments can forbid all cubic terms (e.g. swapping the $C_{18}$ charges of $e^c$ with $\nu^c$ and of $\chi_e$ with $\chi_\nu$ easily achieves this). In a more complete model the auxiliary symmetries and field content (such as an additional familon) would have to be considered in greater detail.

If one were to lift the constraint of a single FSIC, exact alignment would be lost. By keeping the Higgs doublets as FS singlets and generalising from having a single FSIC to a case where any additional FSICs are hierarchically suppressed with respect to a single dominant FSIC, the FS is still responsible for approximately aligning the MHDM Yukawa structures, up to the level of the corrections introduced by the subdominant FSICs.

\section{Conclusion}

We have required exact Yukawa alignment in multi-Higgs doublet models and achieved it by employing a strategy where each family must have only a single allowed family symmetry invariant combination - this is extremely constraining. The models presented are not complete models and are intended as a demonstration of the use of family symmetries in achieving Yukawa alignment. The family symmetries used do not address the flavour issues of the Standard Model at all (e.g. the first model presented is actually unable to accommodate the Standard Model fermion masses). The explanation of the Yukawa couplings is at best shifted into an explanation of the respective familon vacuum expectation values, with the vacuum alignment not being considered here.

It is likely that a better approach to these flavour problems involves abandoning the requirement of a single family symmetry invariant combination. Lifting this constraint makes it easier to construct more realistic examples where the family symmetry also addresses the flavour problems of the Standard Model.
As a generalisation of the models exemplified here we argued that approximate alignment can be achieved if the requirement that the Higgs doublets transform as singlets of the family symmetry is kept, as that enables the family symmetry to construct a similar Yukawa structure for each doublet as long as there is still a single dominant invariant for each family - and this is a natural expectation given the observed hierarchy of fermion masses. The goal in such an approach would be that the approximate alignment sufficiently suppresses the unobserved processes - with the interesting possibility that new physics signals may be just beyond the current experimental reach.

There are also other possibilities to achieve Yukawa alignment in multi-Higgs doublet models that can be explored. An entirely distinct approach lies in having a symmetry that acts on the Higgs doublets, possibly achieving alignment through additional doublets that do not couple directly to the fermions \cite{Hugo}.

\section*{Acknowledgements}

We are grateful to David Emmanuel-Costa for reading the manuscript and to Hugo Ser\^{o}dio for helpful discussions.
This work was supported by DFG grant PA 803/6-1.

\bibliography{refs}

\begin{thebibliography}{15}
\expandafter\ifx\csname natexlab\endcsname\relax\def\natexlab#1{#1}\fi
\expandafter\ifx\csname bibnamefont\endcsname\relax
  \def\bibnamefont#1{#1}\fi
\expandafter\ifx\csname bibfnamefont\endcsname\relax
  \def\bibfnamefont#1{#1}\fi
\expandafter\ifx\csname citenamefont\endcsname\relax
  \def\citenamefont#1{#1}\fi
\expandafter\ifx\csname url\endcsname\relax
  \def\url#1{\texttt{#1}}\fi
\expandafter\ifx\csname urlprefix\endcsname\relax\def\urlprefix{URL }\fi
\providecommand{\bibinfo}[2]{#2}
\providecommand{\eprint}[2][]{\url{#2}}

\bibitem[{\citenamefont{Gunion et~al.}(1990)\citenamefont{Gunion, Haber, Kane,
  and Dawson}}]{Haber}
\bibinfo{author}{\bibfnamefont{J.}~\bibnamefont{Gunion}},
  \bibinfo{author}{\bibfnamefont{H.}~\bibnamefont{Haber}},
  \bibinfo{author}{\bibfnamefont{G.}~\bibnamefont{Kane}}, \bibnamefont{and}
  \bibinfo{author}{\bibfnamefont{S.}~\bibnamefont{Dawson}},
  \emph{\bibinfo{title}{The Higgs Hunter's Guide}}
  (\bibinfo{publisher}{Addison-Wesley}, \bibinfo{address}{New York},
  \bibinfo{year}{1990}).

\bibitem[{\citenamefont{Branco et~al.}(1999)\citenamefont{Branco, Lavoura, and
  Silva}}]{Gustavo}
\bibinfo{author}{\bibfnamefont{G.~C.} \bibnamefont{Branco}},
  \bibinfo{author}{\bibfnamefont{L.}~\bibnamefont{Lavoura}}, \bibnamefont{and}
  \bibinfo{author}{\bibfnamefont{J.~P.} \bibnamefont{Silva}},
  \emph{\bibinfo{title}{CP Violation}} (\bibinfo{publisher}{Oxford University
  Press}, \bibinfo{address}{Oxford}, \bibinfo{year}{1999}).

\bibitem[{\citenamefont{Pich and Tuzon}(2009)}]{PT_a}
\bibinfo{author}{\bibfnamefont{A.}~\bibnamefont{Pich}} \bibnamefont{and}
  \bibinfo{author}{\bibfnamefont{P.}~\bibnamefont{Tuzon}},
  \bibinfo{journal}{Phys.Rev.} \textbf{\bibinfo{volume}{D80}},
  \bibinfo{pages}{091702} (\bibinfo{year}{2009}), \eprint{0908.1554}.

\bibitem[{\citenamefont{Ferreira et~al.}(2010)\citenamefont{Ferreira, Lavoura,
  and Silva}}]{Lavoura_RG}
\bibinfo{author}{\bibfnamefont{P.}~\bibnamefont{Ferreira}},
  \bibinfo{author}{\bibfnamefont{L.}~\bibnamefont{Lavoura}}, \bibnamefont{and}
  \bibinfo{author}{\bibfnamefont{J.~P.} \bibnamefont{Silva}},
  \bibinfo{journal}{Phys.Lett.} \textbf{\bibinfo{volume}{B688}},
  \bibinfo{pages}{341} (\bibinfo{year}{2010}), \eprint{1001.2561}.

\bibitem[{\citenamefont{Braeuninger et~al.}(2010)\citenamefont{Braeuninger,
  Ibarra, and Simonetto}}]{Ibarra_RG}
\bibinfo{author}{\bibfnamefont{C.~B.} \bibnamefont{Braeuninger}},
  \bibinfo{author}{\bibfnamefont{A.}~\bibnamefont{Ibarra}}, \bibnamefont{and}
  \bibinfo{author}{\bibfnamefont{C.}~\bibnamefont{Simonetto}},
  \bibinfo{journal}{Phys.Lett.} \textbf{\bibinfo{volume}{B692}},
  \bibinfo{pages}{189} (\bibinfo{year}{2010}), \eprint{1005.5706}.

\bibitem[{\citenamefont{Ross et~al.}(2004)\citenamefont{Ross, Velasco-Sevilla,
  and Vives}}]{Oscar1}
\bibinfo{author}{\bibfnamefont{G.~G.} \bibnamefont{Ross}},
  \bibinfo{author}{\bibfnamefont{L.}~\bibnamefont{Velasco-Sevilla}},
  \bibnamefont{and} \bibinfo{author}{\bibfnamefont{O.}~\bibnamefont{Vives}},
  \bibinfo{journal}{Nucl.Phys.} \textbf{\bibinfo{volume}{B692}},
  \bibinfo{pages}{50} (\bibinfo{year}{2004}), \eprint{hep-ph/0401064}.

\bibitem[{\citenamefont{Calibbi et~al.}(2010)\citenamefont{Calibbi,
  Jones-Perez, Masiero, Park, Porod et~al.}}]{Oscar2}
\bibinfo{author}{\bibfnamefont{L.}~\bibnamefont{Calibbi}},
  \bibinfo{author}{\bibfnamefont{J.}~\bibnamefont{Jones-Perez}},
  \bibinfo{author}{\bibfnamefont{A.}~\bibnamefont{Masiero}},
  \bibinfo{author}{\bibfnamefont{J.-h.} \bibnamefont{Park}},
  \bibinfo{author}{\bibfnamefont{W.}~\bibnamefont{Porod}},
  \bibnamefont{et~al.}, \bibinfo{journal}{Nucl.Phys.}
  \textbf{\bibinfo{volume}{B831}}, \bibinfo{pages}{26} (\bibinfo{year}{2010}),
  \eprint{0907.4069}.

\bibitem[{\citenamefont{Lalak et~al.}(2010)\citenamefont{Lalak, Pokorski, and
  Ross}}]{Graham}
\bibinfo{author}{\bibfnamefont{Z.}~\bibnamefont{Lalak}},
  \bibinfo{author}{\bibfnamefont{S.}~\bibnamefont{Pokorski}}, \bibnamefont{and}
  \bibinfo{author}{\bibfnamefont{G.~G.} \bibnamefont{Ross}},
  \bibinfo{journal}{JHEP} \textbf{\bibinfo{volume}{1008}}, \bibinfo{pages}{129}
  (\bibinfo{year}{2010}), \eprint{1006.2375}.

\bibitem[{\citenamefont{de~Medeiros~Varzielas and Ross}(2006)}]{Ivo1}
\bibinfo{author}{\bibfnamefont{I.}~\bibnamefont{de~Medeiros~Varzielas}}
  \bibnamefont{and} \bibinfo{author}{\bibfnamefont{G.~G.} \bibnamefont{Ross}},
  \bibinfo{journal}{Nucl.Phys.} \textbf{\bibinfo{volume}{B733}},
  \bibinfo{pages}{31} (\bibinfo{year}{2006}), \eprint{hep-ph/0507176}.

\bibitem[{\citenamefont{de~Medeiros~Varzielas}(2008)}]{Ivo2}
\bibinfo{author}{\bibfnamefont{I.}~\bibnamefont{de~Medeiros~Varzielas}}
  (\bibinfo{year}{2008}), \eprint{0804.0015}.

\bibitem[{\citenamefont{de~Medeiros~Varzielas and Merlo}(2011)}]{IvoLuca}
\bibinfo{author}{\bibfnamefont{I.}~\bibnamefont{de~Medeiros~Varzielas}}
  \bibnamefont{and} \bibinfo{author}{\bibfnamefont{L.}~\bibnamefont{Merlo}},
  \bibinfo{journal}{JHEP} \textbf{\bibinfo{volume}{1102}}, \bibinfo{pages}{062}
  (\bibinfo{year}{2011}), \eprint{1011.6662}.

\bibitem[{\citenamefont{Luhn et~al.}(2007)\citenamefont{Luhn, Nasri, and
  Ramond}}]{Luhn1}
\bibinfo{author}{\bibfnamefont{C.}~\bibnamefont{Luhn}},
  \bibinfo{author}{\bibfnamefont{S.}~\bibnamefont{Nasri}}, \bibnamefont{and}
  \bibinfo{author}{\bibfnamefont{P.}~\bibnamefont{Ramond}},
  \bibinfo{journal}{J.Math.Phys.} \textbf{\bibinfo{volume}{48}},
  \bibinfo{pages}{123519} (\bibinfo{year}{2007}), \eprint{0709.1447}.

\bibitem[{\citenamefont{King and Luhn}(2009)}]{Steve1}
\bibinfo{author}{\bibfnamefont{S.~F.} \bibnamefont{King}} \bibnamefont{and}
  \bibinfo{author}{\bibfnamefont{C.}~\bibnamefont{Luhn}},
  \bibinfo{journal}{Nucl.Phys.} \textbf{\bibinfo{volume}{B820}},
  \bibinfo{pages}{269} (\bibinfo{year}{2009}), \eprint{0905.1686}.

\bibitem[{\citenamefont{King and Luhn}(2010)}]{Steve2}
\bibinfo{author}{\bibfnamefont{S.~F.} \bibnamefont{King}} \bibnamefont{and}
  \bibinfo{author}{\bibfnamefont{C.}~\bibnamefont{Luhn}},
  \bibinfo{journal}{Nucl.Phys.} \textbf{\bibinfo{volume}{B832}},
  \bibinfo{pages}{414} (\bibinfo{year}{2010}), \eprint{0912.1344}.

\bibitem[{\citenamefont{Ser\^{o}dio}()}]{Hugo}
\bibinfo{author}{\bibfnamefont{H.}~\bibnamefont{Ser\^{o}dio}},
  \bibinfo{note}{private communication}.

\end{thebibliography}

\end{document}